\documentclass[aps,twocolumn,prd,showpacs,showkeys,preprintnumbers,superscriptaddress,nobibnotes,floatfix,longbibliography,nofootinbib]{revtex4-1}

\pdfoutput=1
\usepackage{amsmath}
\usepackage{amsfonts}
\usepackage{amssymb}
\usepackage{mathrsfs}
\usepackage{graphicx}
\usepackage{color}
\usepackage{longtable}
\usepackage{hyperref}
\usepackage{multirow}
\usepackage{slashed}

\newcommand{\pmiss}{\slashed{p}^\perp}

\begin{document}

\title{Discovery and spectroscopy of dark matter and dark sectors \\ with microscopic black holes at next generation colliders}

\author{Ningqiang Song}
\email{ningqiang.song@queensu.ca}

\affiliation{Arthur B. McDonald Canadian Astroparticle Physics Research Institute, Department of Physics, Engineering Physics and Astronomy, Queen's University, Kingston ON K7L 3N6, Canada}
\affiliation{Perimeter Institute for Theoretical Physics, Waterloo ON N2L 2Y5, Canada}

\author{Aaron C. Vincent}
\email{aaron.vincent@queensu.ca}
\affiliation{Arthur B. McDonald Canadian Astroparticle Physics Research Institute, Department of Physics, Engineering Physics and Astronomy, Queen's University, Kingston ON K7L 3N6, Canada}
\affiliation{Perimeter Institute for Theoretical Physics, Waterloo ON N2L 2Y5, Canada}


\begin{abstract}
If the length scale of possible extra dimensions is large enough, the effective Planck scale is lowered such that microscopic black holes could be produced in collisions of high-energy particles at colliders. These black holes evaporate through Hawking radiation of a handful of energetic particles drawn from the set of all kinematically and thermally allowed degrees of freedom, including dark matter.  Here, we perform the first numerical black hole spectroscopic study of the dark sector. We find that if the next generation of colliders can produce microscopic black holes, then missing momentum signatures can reveal the existence of \textit{any new light ($\lesssim 10$ {\rm TeV}) particle}, regardless of the strength of its coupling to the Standard Model, even if there exists no such non-gravitational coupling at all.
\end{abstract}

\maketitle

\textbf{\textit{Introduction --- }} Significant experimental effort is underway to uncover the nature of Dark Matter (DM), which makes up approximately 85\% of the matter in the observable Universe \cite{Aghanim:2018eyx}. Cosmological observations dictate that the DM cannot interact strongly with particles from the Standard Model (SM) of particle physics, and must behave on large scales like a cold, collisionless particle. There are theoretical motivations for a particle physics ``portal'' to the dark sector (DS) via some weak-scale interaction, that mainly have to do with the relative abundance of DM versus SM matter in the Universe \cite{Scherrer:1985zt,Kaplan:2009ag}. Despite extensive experimental searches, all evidence of DM has remained in its gravitational interactions on the scale of galaxies and larger. 

There exists a possibility, sometimes dubbed the ``nightmare scenario'', that the DM --- and indeed its possible extension to the dark sector --- is entirely secluded from the SM, interacting only via gravity. This is a disheartening prospect. However, we will show that in the presence of large enough extra dimensions to bring the Planck scale down near the electroweak scale, all hope is not lost.

Large Extra Dimensions (LEDs) have been proposed for a multitude of theoretical and phenomenological reasons, and TeV scale quantum gravity in particular offers a compelling solution to the hierarchy problem \cite{ArkaniHamed:1998rs,Antoniadis:1998ig,ArkaniHamed:1998nn,Argyres:1998qn,Randall:1999ee,Randall:1999vf}  that exists between the Planck and electroweak scales. In such scenarios, the SM is confined to a 3+1 dimensional \textit{brane}, and only gravity may propagate in the full $D = 4+n$ dimensional \textit{bulk}. Depending on the number, size and geometry of the LEDs, the true Planck scale $M_\star$ can  be much lower than the effective Planck scale $M_{Pl} \sim 10^{18}$ GeV seen on our brane. These scenarios have been tested extensively in terms of gravitational force \cite{Murata:2014nra}, supernova and neutron star cooling \cite{Hanhart:2001fx,Hannestad:2003yd}, the metastability of the Higgs vacuum \cite{Mack:2018fny}, and at colliders \cite{Sirunyan:2018xwt,Sirunyan:2018wcm}, as we shall scrutinize below.  If only one LED exists, this solution leads to modifications of gravity on Solar System scales \cite{ArkaniHamed:1998rs}. However, two or more LEDs remain allowed. Current constraints limit the scale $M_\star$ to be above about 3-25 TeV \cite{Murata:2014nra,Hanhart:2001fx,Hannestad:2003yd,Mack:2018fny,Sirunyan:2018xwt,Sirunyan:2018wcm}. 

A profound consequence of a low Planck scale is the possibility of creating microscopic black holes (BHs) in high-energy collisions. Indeed, the hoop conjecture \cite{thorne1995black,Banks:1999gd} implies that as long as the impact parameter $b$ between two colliding particles is smaller than twice the horizon radius $r_h$ of a BH with mass $M_\bullet = E_{CM}$, the centre of mass energy, then a BH will form. The phenomenology of BHs at colliders is a rich and well-studied topic \cite{Dimopoulos:2001hw,Giddings:2001bu,Chamblin:2004zg,Harris:2004mf,Cavaglia:2006uk,Alberghi:2006km,Cavaglia:2007ir,Calmet:2008dg,Nayak:2009fv,Gingrich:2009hj,Gingrich:2009da,Landsberg:2014bya}, and has mainly focused on the prospects and methods of detecting such objects, as they rapidly evaporate via Hawking radiation \cite{Hawking:1974sw}. 

BH evaporation is effectively instantaneous, and because of the very small masses, the BH will only emit a few ($\sim$ 6-20, depending on initial BH mass) particles in its decay. Crucially, decay products are a  subset of \textit{every} bulk degree of freedom, drawn from a thermal distribution at the Hawking temperature $T_H$, independently of any particle physics. By carefully measuring the energy and momentum output of a large enough sample of decaying BHs, one in principle has full access to any sequestered DS. In practice, missing energy and momentum during a decay process allow us to count the number of invisible degrees of freedom, including neutrinos, gravitons and dark matter, allowing for a discovery of the DS particles. We also note that the aforementioned collider bounds on LEDs rely on the assumption that BHs only decay to SM matter and gravitions. Evaporation into a large DS may lead to a reconsideration of these bounds, and the Planck scale can be significantly lowered.

Two basic conditions need to be fulfilled: 1) the full $D$-dimensional Planck scale must be low enough to allow production at future colliders, and 2) the masses of the DS particles must be lower than the BH mass. As long as $T_H$, which is typically large, remains near or above the DM mass, production is not Boltzmann-suppressed.  There is no lower limit to the masses that can be probed. For this study, we consider a dark sector with particle masses $m_{i} \ll M_\star$. Our conclusions hold in the most generic scenario in which all particles except gravitons are confined to our 3-brane, or to one that is parallel. Even if there is a dark sector that can propagate in the bulk, only the projection of its momentum onto our brane can be seen. However, emission in the bulk is kinematically suppressed because the BH must conserve gauge charge, and will thus be confined to our brane in most instances.

 Some of the strongest limits on BH production come from the Large Hadron Collider, with CM collisions limited to 14 TeV. We thus turn to the next generation of colliders, such as the Future Circular Collider (FCC) \cite{Benedikt:2015kqj} or Super Proton-Proton Collider (SPPC) \cite{Tang:2015qga}, with a CM energy of 100 TeV. 

We consider a simple LED scenario, with a single, tensionless brane, leaving a fuller exploration of the LED parameter space to future work. We will assume that the actual detection of such BHs is well-understood, and focus on the use of BHs as probes of new particle physics. This is  reasonable, as the high-multiplicity, high-transverse momentum signature of a BH event is very distinct from known SM processes, appearing in signal regions with almost no expected SM background (see e.g. \cite{Chatrchyan:2013xva,Aad:2013lna} and references above).

We first review the essential elements of BH production and evaporation at colliders. Then, we incorporate the extended DS into a full simulation of BH production and evaporation in order to carefully quantify the missing momentum signature accessible to future colliders. We end with a discussion of future directions.

\textbf{\textit{BH production and decay at colliders ---}}
The BH production cross section is \cite{Dai:2007ki}:
\begin{equation}
    \sigma^{pp\rightarrow BH} = \int_{M_\star^2/s}^1 du \int_u^1 \frac{dv}{v}\pi b_{\mathrm{max}}^2 \sum_{i,j}f_i(v,Q)f_j(u/v,Q)
    \label{eq:xsec}
\end{equation}
where\footnote{we work in natural units: $c = \hbar = k_B = 1$.} $b_{\mathrm{max}} = 2r_s^{(D)}(E_{CM})/[1+(D-2)^2/4]^{1/(D-3)}$ is the maximum impact parameter that allows BH production, $\sqrt{s}$ is the center of mass energy of a hadron-hadron collision and $ $
\begin{equation}
    r_s^{(D)} = k_D M_\star^{-1}\left(\frac{\sqrt{us}}{M_\star} \right)^{1/(D-3)}
\end{equation}
is the $D$-dimensional Schwarzschild radius of a BH with mass $M_\bullet = \sqrt{us} \equiv E_{CM}$. $k_D$ is a geometrical factor related to the number of extra dimensions. The factors of $f(v,Q)$ represents the parton distribution functions (PDF) of each constituent of the colliding hadrons, which are summed over all quark flavors and gluons. $v$ and $u/v$ are the fraction of CM energy carried by each colliding parton, and $Q^2$ is the square of the exchanged momentum four-vector.  The formation of a stationary BH is usually associated with partial loss of its energy, momentum and angular momentum due to the nonlinearity in the collision, which can affect the total formation cross section. We do not include this initial energy loss, as only a few numerical studies have attempted to quantify it \cite{Yoshino:2005hi}. Because the semiclassical treatment \eqref{eq:xsec} must break down near $M_\star$, the minimum energy required to form a BH, $M_{\rm min}$ could be different. In the absence of a detailed theory, we make the simplifying assumption $M_{\rm min} = M_\star$.
\begin{figure*}[ht]
    \centering
        \begin{tabular}{c c}
    \includegraphics[width = 0.5\textwidth]{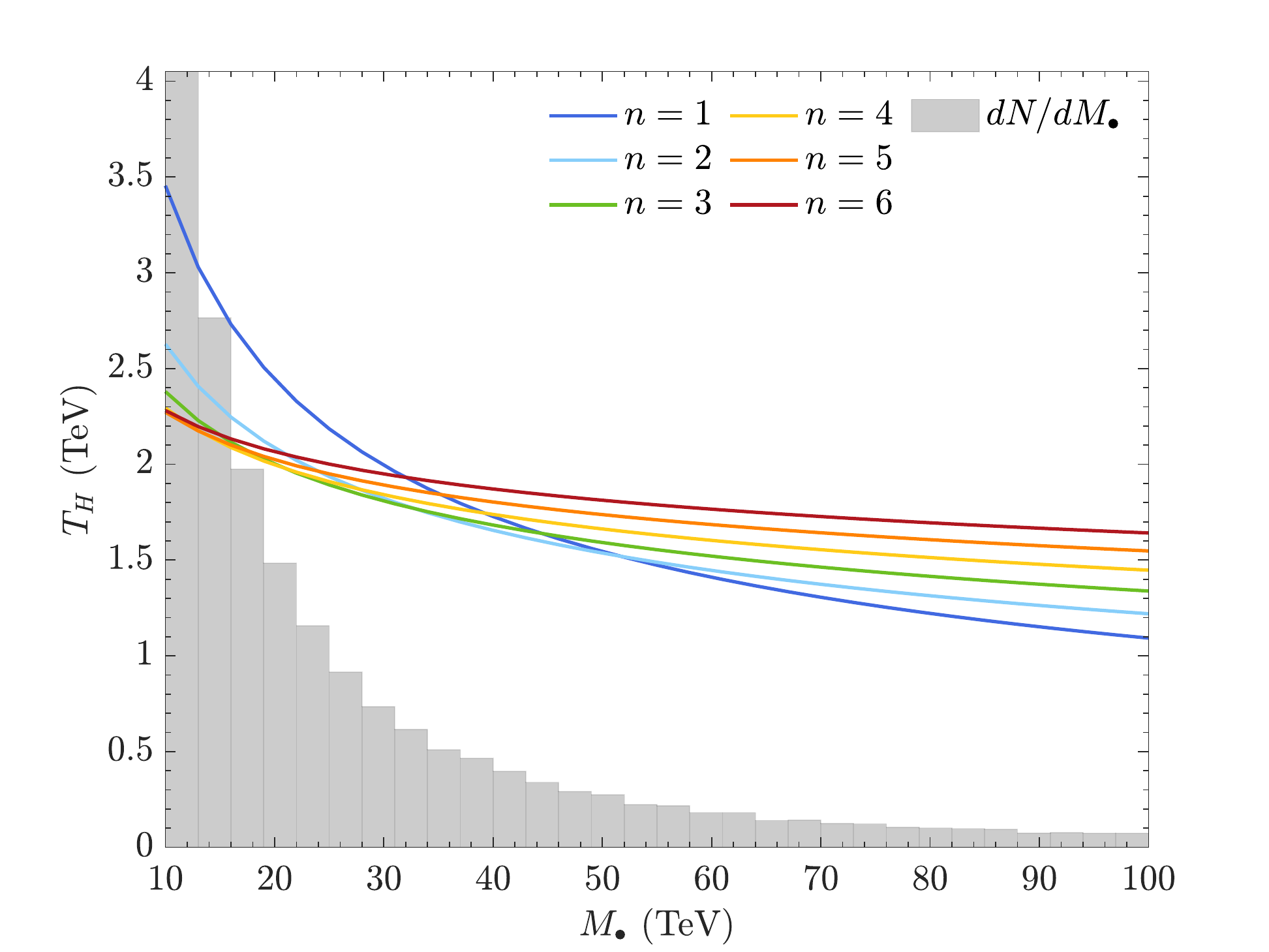} &
        \hspace{-.8cm}\includegraphics[width = 0.55\textwidth]{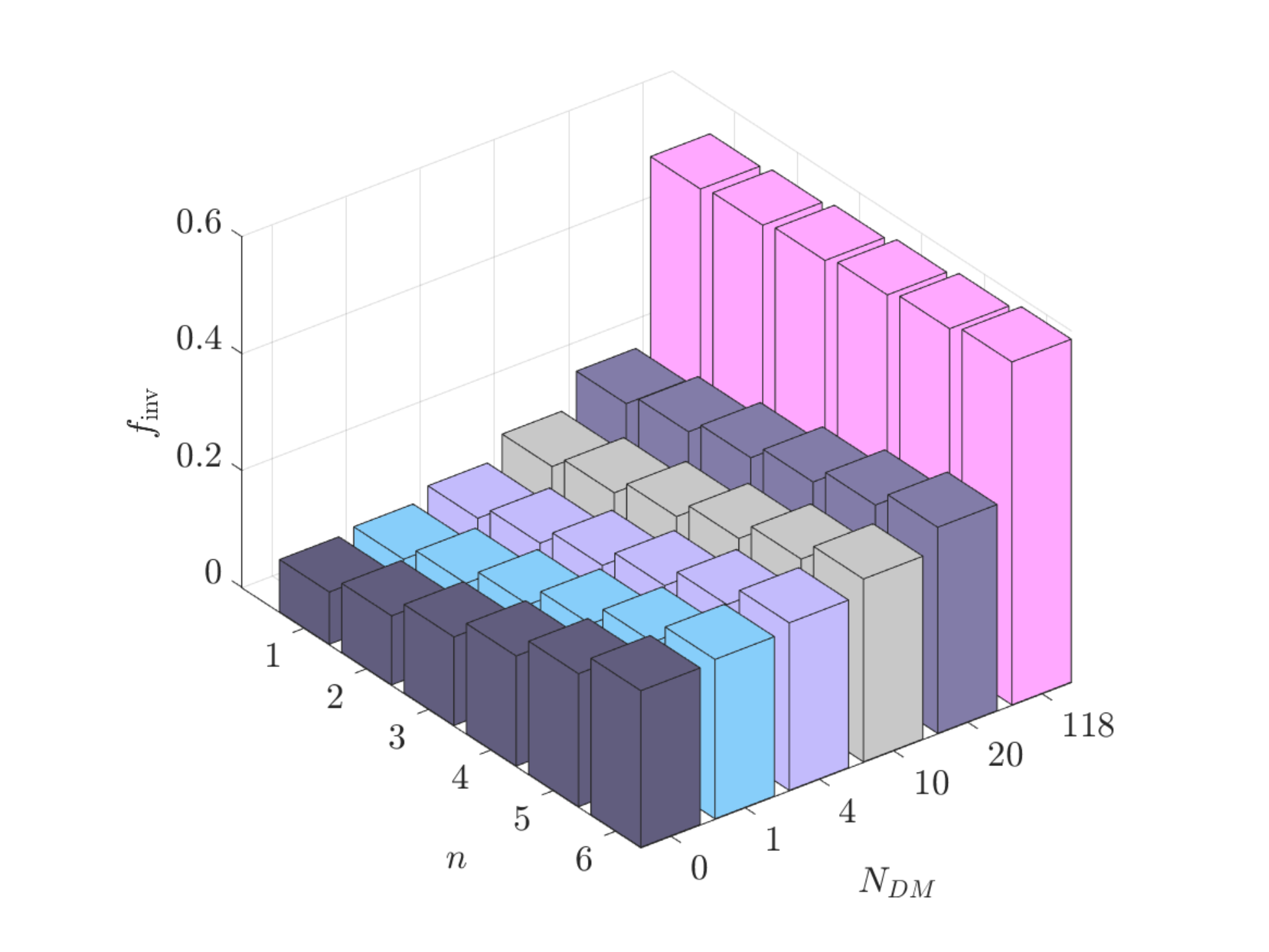}  \\
    \end{tabular}
    \caption{\textbf{Left:} Lines: Hawking Temperature $T_H$ as a function of the microscopic BH mass, for $M_\star = 10$ TeV in the case of $n = $ 1 to 6 extra dimensions. Grey bands represent the distribution of BH masses created in 100 TeV proton-proton collisions (arbitrary units). \textbf{Right:} the fraction of invisible energy released from the evaporation of BHs, for different values of the number of extra dimensions $n$, and dark sector degrees of freedom, $N_{DM}$. }
    \label{fig:fmiss}
\end{figure*}

Once produced, the BH will evaporate via Hawking radiation, followed by what is modeled as a final ``burst'' as the BH mass approaches $M_\star$.  Hawking evaporation proceeds with a temperature $T_H = (D-3)/4\pi r_h$ (shown in the left panel of Fig.~\ref{fig:fmiss}, see later for detailed discussion). In the absence of angular momentum, the horizon radius $r_h$ is equal to $r_s$. There are two contributions to an effective deviation from a thermal spectrum: first, a ``greybody'' factor must be included to account for the distortion of the thermal spectrum as it is ultimately seen by an observer at ``infinity'', rather than at the horizon. Second, since $T_H$ is similar in magnitude to $M_\bullet$, only a few particles are produced per BH, so a continuum is not expected. Moreover, as a complete theory of gravity below $M_\star$ is missing, the semiclassical approach is bound to fail in the final stages of evaporation. We make the typical assumption that a minimum number of particles that conserve the energy, momentum, angular momentum and all the gauge charges of the BH are emitted in a final burst when $M_\bullet$ nears $M_\star$.

Including all polarization states, the SM contains 118 degrees of freedom (DoF), of which 6 (neutrinos) can be considered completely invisible. A BH may also decay invisibly to gravitons, which have $D(D-3)/2$ polarization states.

The fraction of the BH energy emitted invisibly should be approximately 
\begin{equation}
    f_{\mathrm{inv}} = \frac{N_{\nu} + N_G + N_{DM}}{N_{vis} + N_{\nu} + N_G + N_{DM}},
    \label{eq:finv}
\end{equation}
 where $N_{vis}$, $N_{\nu}$, $N_G$, and $N_{DM}$ are the respective number of DoFs in the visible, neutrino, graviton and dark sectors. The fraction of invisible decay products from evaporating BHs with one extra dimension is around 9\% when only SM particles are present; this rises to 12\% in the presence of a single dark Dirac fermion (4 DoF), a $\sim 30\%$ increase. Fig. \ref{fig:fmiss} shows this fraction as a function of the number of extra dimensions $n = D -4$ and the number of dark degrees of freedom $N_{DM}$.
 
Heavy leptons, weak bosons and hadronic jets further release some energy into neutrinos, and other charged decay products may have too low an energy to be detected. We account for the former in our full simulations, but not the latter as we are interested in the theoretical expectation. 

In a hadron collider, collision  occurs between two partons, as in Eq.~\eqref{eq:xsec}, so the total energy in the interaction is not known. We thus rely on missing transverse momentum, denoted $\pmiss$, defined as the total unbalanced momentum transverse to the collision axis. To compute missing momenta, we turn to a full numerical simulation of BH production, evaporation and showering.

\textbf{\textit{Numerical simulations --- }}
Because of the small (20 or fewer) number of particles produced per BH evaporation, collider signatures can be highly anisotropic. To simulate this process, we first employ a modified version of the BlackMax \cite{Dai:2007ki,Dai:2009by} code. BlackMax is a Monte Carlo event generator which simulates production and evaporation of BHs in $pp$, $p \bar p$ and $e^+ e^-$ collisions. It includes known greybody factors, can account for nonzero angular momentum, and BH recoil in the bulk. Given a fixed proton-proton CM energy, BlackMax generates black holes distributed as an unintegrated Eq. \eqref{eq:xsec}, i.e. proportional to $d \sigma^{pp\rightarrow BH} /du$.

Rather than using the built-in PYTHIA 6 interface, we pass the results of BlackMax to PYTHIA 8 \cite{Sjostrand:2014zea} to account for QCD radiation, hadronization of quark and gluon final states and decay of heavy particles. In both steps, we employ the CT14NNLO \cite{Dulat:2015mca} PDFs, implemented with LHAPDF6 \cite{Buckley:2014ana}.

We modify the BlackMax code to add a variable number of new dark particle DoFs to the BH evaporation spectrum. We take each DoF to obey the thermal and greybody distributions of scalar bosons\footnote{We have verified that using Fermi-Dirac statistics and greybody factors instead does not significantly alter our conclusions.}. Greybody factors for bulk emission of gravitons from rotating BHs are not available (specifically superradiance can be important for extremely rotating BHs~\cite{Stojkovic:2004hp}), and are thus not implemented in BlackMax. We therefore simplify our simulations to consider only nonrotating BHs, since the emission of bulk gravitons has a more important effect on missing energy. 

We simulate BH production and evaporation in the case of 1 to 6 extra dimensions.\footnote{$n = 1$ is strongly excluded, but we keep it for illustrative purposes.} We take the proton-proton CM energy $\sqrt{s} = 100$ TeV, and the Planck scale $M_\star = 10$ TeV, to be compatible with current collider multijet \cite{Sirunyan:2018xwt} and dijet \cite{Sirunyan:2018wcm} searches. We also assume the minimum BH mass $M_\mathrm{min}$ is of the order of $M_\star$ and neglect any possible energy loss in the phase before the formation of a stationary BH.

For each scenario, we simulate 10$^4$ BH events, assuming 0, 1, 4, 10, 20 and 118 dark DoFs, as a representation of possible extended dark sectors. 4 DoFs could represent a single Dirac fermion; O(10) DoFs could include new dark forces, while 118 DoFs correspond to an entire ``mirror'' sector of the SM \cite{Foot:1991bp}. 

We assume that BH events can be clearly identified, and focus on what we learn from missing transverse momentum from such events. After primary particle hadronization and decay, the momenta of the remaining visible particles are added vectorially, in the plane transverse to the beam pipe. For each BH, the magnitude of this vector corresponds by definition to $\pmiss$. 

\begin{figure}
  \hspace*{-.7cm} \includegraphics[trim= 0 0 0 0,clip,width = 0.55\textwidth]{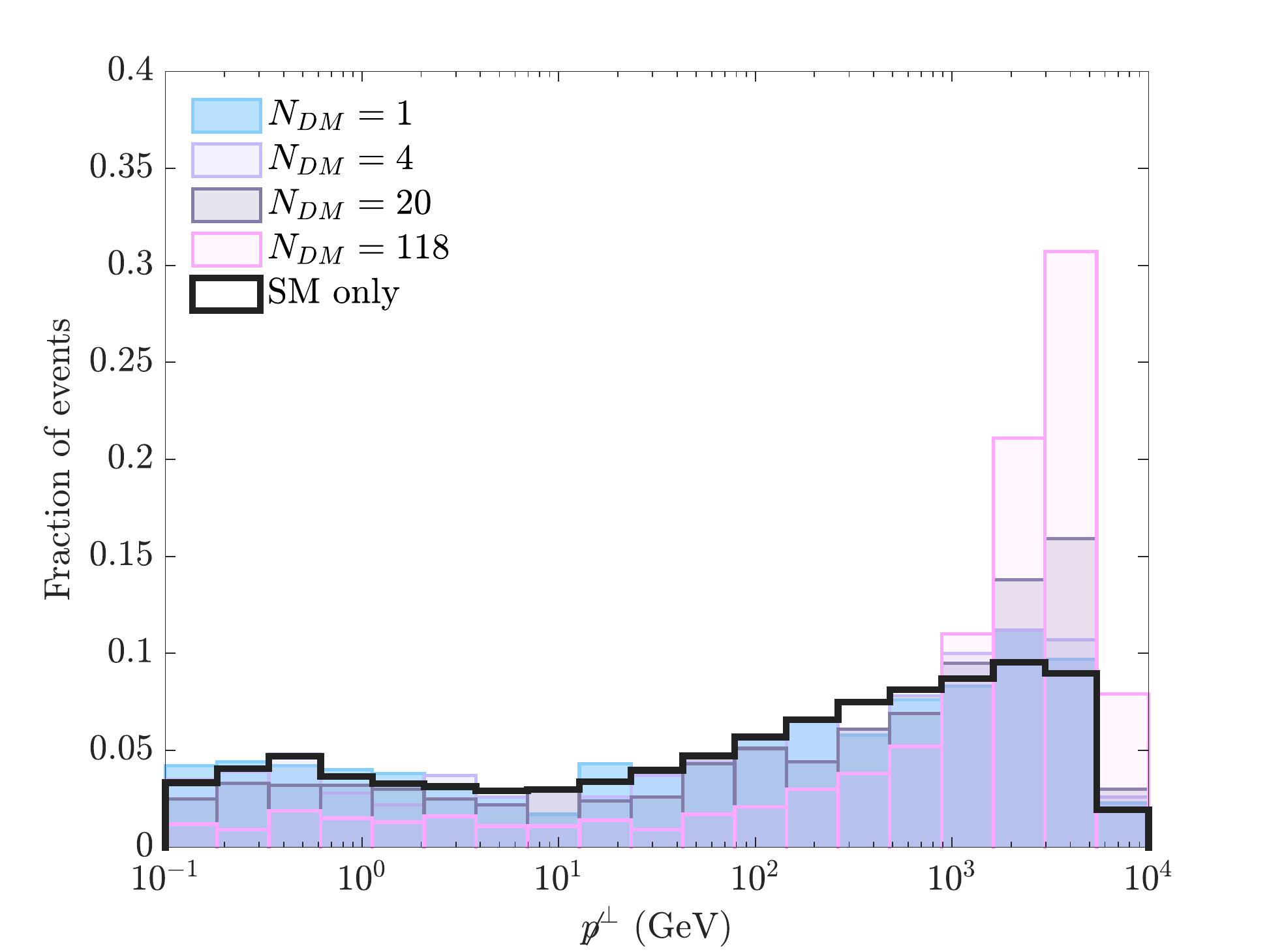}
    \caption{Distribution of missing transverse momentum from BH evaporation for $n =2$ extra dimensions. In black we show the SM-only case, and in colour, the case where the dark sector contains $N_{DM} = $ 1, 4, 20 and 118 new particles with masses $m_{i} < M_\bullet$. Missing momentum from SM background sources are not included here.}
    \label{fig:pmiss2ed}
\end{figure}

Fig. \ref{fig:pmiss2ed} is a histogram of $\pmiss$ from a sample of simulated evaporating BHs in the case of two extra dimensions, when only neutrinos and gravitons are emitted (black, $N_{DM} = 0$), and for $N_{DM} =$ 1, 4, 20 and 118 new DoFs. As the number of dark DoFs increase, the mean $\pmiss$ rises sharply, and the distribution becomes much more peaked. This is due to the increase in primary particles which carry away a large fraction of the BH's energy. BHs decaying to small amounts of $\pmiss$, which mainly comes from the partial invisible decay of heavy SM states, become rarer. This trend does not change with the number of extra dimensions. We note that the main difference in distributions comes from the high $\pmiss$ region, further motivating our decision to neglect low-energy cuts. As stated earlier, we are not including backgrounds from SM processes (i.e. when no BH is formed), as they are expected to be sufficiently far from the BH signal region.

\begin{figure}
   \hspace*{-.7cm}\includegraphics[trim= 0 0 0 0,clip,width = 0.55\textwidth]{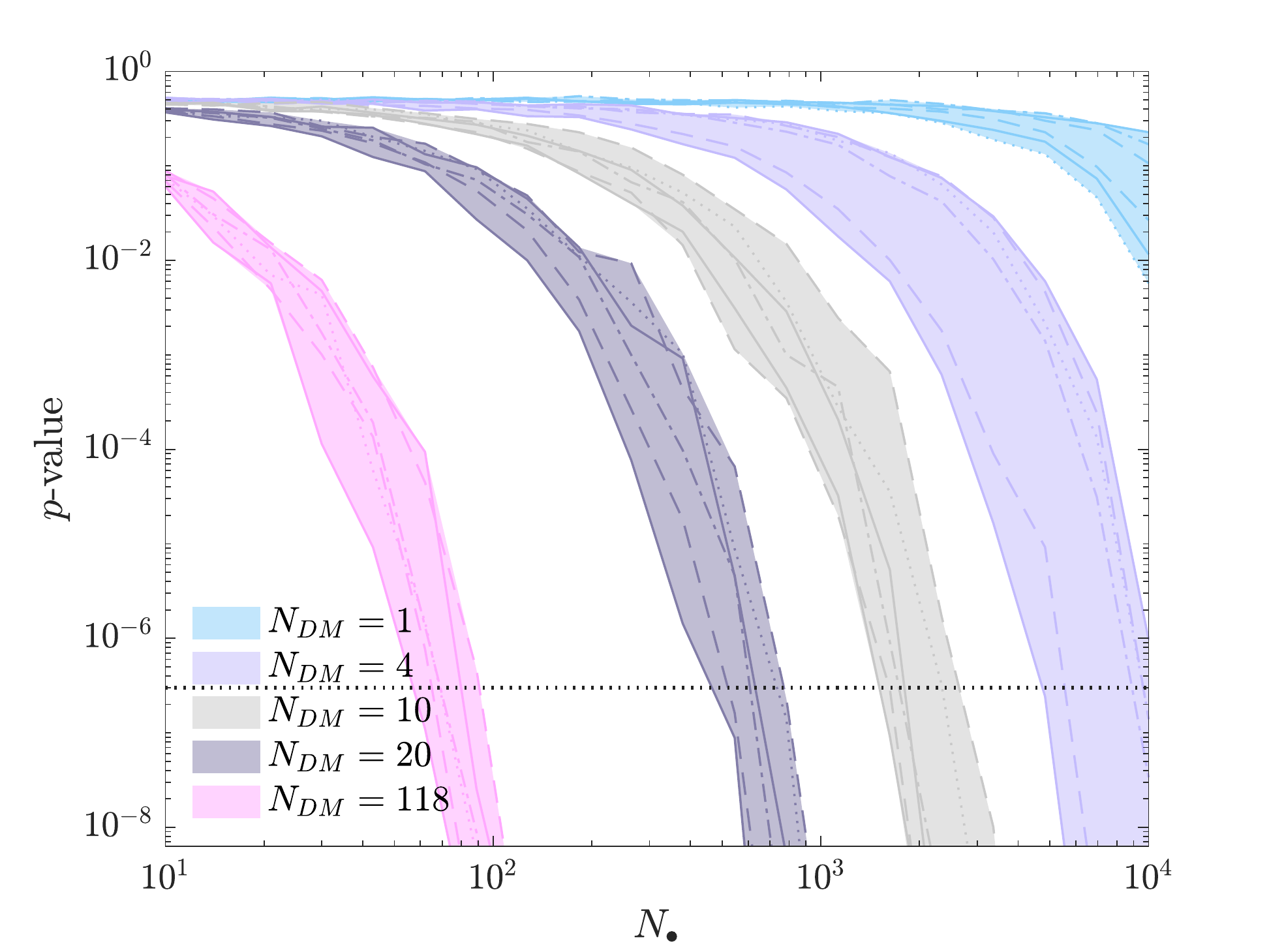}
    \caption{Kolmogorov-Smirnov $p$-values obtained from missing transverse momentum observations, due to the evaporation of microscopic BHs into dark sector particles, compared with the case where only SM particles exist. We assume the number of extra dimensions is known. Different light bands represent dark sectors with 1 (blue, top), 4 (purple), 10 (light grey) 20 (dark grey) and 118 (pink, bottom-left) new particles.  Within each band, thin lines are the specific cases of 1 (leftmost),  2, 3, 4, 5 and 6 (rightmost, least constrained) large extra dimensions. The horizontal dotted line depicts the 5$\sigma$ significance threshold.}
    \label{fig:KSpvalues}
\end{figure}

Assuming the missing momentum for each event can be well-reconstructed, we wish to know how many events must be observed to establish the existence of an extended DS. We construct reference distributions for the cases of $n=1$ to $6$ extra dimensions using 10$^6$ SM-only (i.e. $N_{DM}=0$)  simulated events each (for a total of $\sim$ 150 CPU days), which we use to compute a Kolmogorov-Smirnov (KS) $p$-value, where we have made the assumption that the number of LEDs is known.  The KS test assesses how likely it is that a given data sample was drawn from the reference distribution, by comparing the maximum difference in their cumulative distribution functions, and is independent of any binning. The interpretation of the $p$-value remains the same as other methods.

Fig. \ref{fig:KSpvalues} shows the KS $p$-values, depending on the number of BH events recorded $N_\bullet$, which we sample randomly from our simulated events. For each value of $N_\bullet$ we draw 300 samples, and average the $p$-value to obtain an expected sensitivity. Small features nonetheless remain, reflecting the random sampling. We have verified that adding a 10\% random Gaussian error on the reconstructed momenta does not affect our results, and the main source of uncertainty remains statistical. The coloured bands show the full range of $p$-values spanned by the six cases of $n = $ 1--6 extra dimensions. The thin lines show these results for each value of $n$, where $n = 1$ is always the strongest-constrained (leftmost), and $n = 6$ is the weakest. This is as we expect from Fig.~\ref{fig:fmiss}. Most BHs produced are near the 10 TeV threshold, where $T_H$ is highest at low $n$, leading to higher-energy particles; in higher dimensions, the extra gravitons add to the background missing energy, while the energy per DM particle is lower.  For reference, $p < 3\times 10^{-7}$, equivalent to the standard 5$\sigma$ discovery threshold, is shown as a horizontal dotted line.

 Depending on the number of extra dimensions, and the size of the extended dark sector, only a few hundred to a few thousand BH events are required to robustly determine its existence. For $N_{DM} = 1$, more than $10^4$ events are required. In the case of 4 DoFs, a strong detection can be made with 5,000 to 10,000 events. Extending the DS improves prospects significantly: if the DS contains as many DoFs as the SM, fewer than 100 events can robustly establish its existence.

As a consistency check, we also performed a binned chi-square test. The significance thus obtained is similar in behaviour to the KS results, though slightly weaker, owing to the KS method's larger sensitivity to deviations from the expected distribution. These results are available in the supplementary material. We furthermore checked that the change in $\pmiss$ distributions follows the same trend when angular momentum is included (also shown in the supplementary material). We did not perform a statistical analysis in this case, because the lack of a bulk graviton emission model for rotating BHs leads to considerable changes in the missing momentum statistics. 

Detecting such an extended dark sector thus requires hundreds (for a large DS) to $\sim 10^4$ observations of BH decay. We find that the cross section to produce BHs for $M_\star = 10$ TeV with a CM beam energy of 100 TeV ranges from 13.5 to 640 picobarns. Producing $N_\bullet$ BH events requires an integrated luminosity that scales as:
\begin{equation}
   \frac{L}{6 \times 10^{36} \, \mathrm{cm^{-2}}} \simeq  \frac{N_\bullet}{1000} 10^{0.061(n-6)^2},
\end{equation}
or between $10^{38}$ cm$^{-2}$ ($n = 2$ extra dimensions) and  $6 \times 10^{36}$ cm$^{-2}$ ($n = 6$) to produce $\sim 1000$ BHs. Estimates of the FCC-hh requirements and capacity put its luminosity at 0.2 to 2 inverse attabarns ($10^{42}$ cm$^{-2}$) \cite{Zimmermann:2016puu}. If LEDs are large enough, a sufficient quantity will be produced in the next generation of hadron colliders to unveil the nature of the DS, as long as its constituent particles are lighter than the BHs being produced.

\textbf{\textit{Discussion and conclusions --- }}
If the fundamental scale of quantum gravity indeed lies just above the reach of the LHC, the discovery of large extra dimensions via the creation of BHs opens up a trove of new physics possibilities. We have illustrated this by quantifying the prospects of discovering a new sequestered sector of particle physics. If dark matter couples only via gravity, the BH portal provides one of the only ways of individually detecting DM particles. Because the collider signatures become significantly different with many new degrees of freedom, we note that existing LHC limits may need to be recomputed; the Planck scale may indeed be closer at hand than previously thought.

We have focused on the prospects of simply detecting an extended dark sector. However, many phenomenological questions remain open, including the full effect of different LED properties, such as brane tension and splitting, compactification scheme, and geometry. We have also assumed that the full LED properties were under control, though we do expect some degeneracies between e.g. bulk graviton and DM production. Full spectroscopy of the dark sector remains a difficult task. If the total missing energy and momentum can be obtained, e.g. at an $e^+e^-$ collider, it is possible to reconstruct the mass and spin statistics of invisible primary evaporation products. Nonetheless, if LEDs are present at the $\sim 10$ TeV scale, they become the only guaranteed probe of any new dark sector below $M \sim M_\star$.

\begin{acknowledgments}
We thank Shivesh Mandalia for his help, as well as Joe Bramante and Carlos Tamarit for valuable input. We thank the anonymous referees for insightful comments and suggestions. This work was supported by the Arthur B. McDonald Canadian Astroparticle Physics Research Institute. Computations were performed at the Centre for Advanced Computing, and supported in part by the Canada Foundation for Innovation. Research at Perimeter Institute is supported by the Government of Canada through the Department of Innovation, Science, and Economic Development, and by the Province of Ontario through the Ministry of Research and Innovation.
\end{acknowledgments}
\bibliographystyle{JHEPmod}
\bibliography{refs,nudm}

\section{Supplementary material}
\begin{figure*}
    \begin{tabular}{c c}
   \includegraphics[trim= 30 0 0 0,clip,width = 0.5\textwidth]{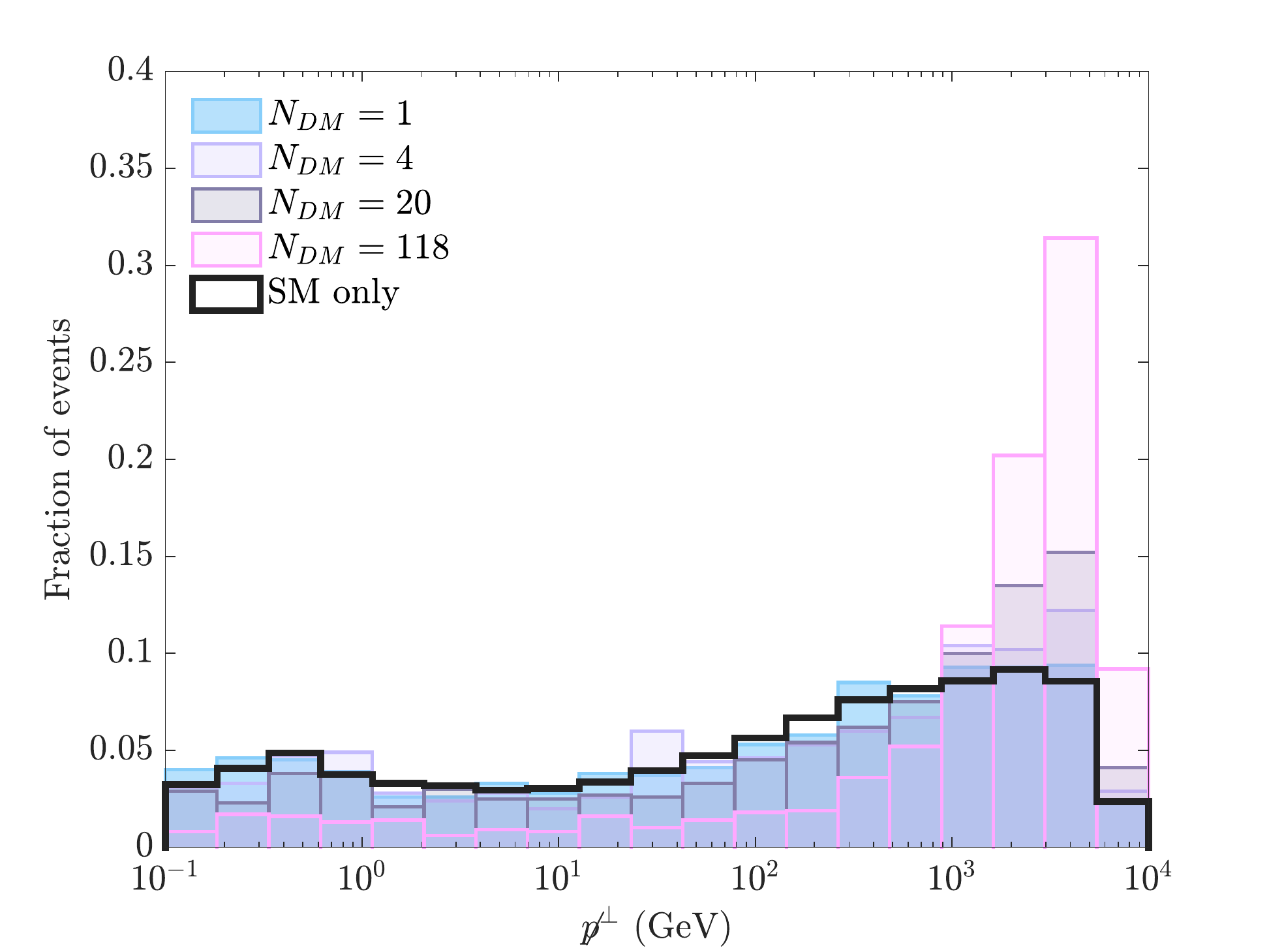} &
      \includegraphics[trim= 30 0 0 0,clip,width = 0.5\textwidth]{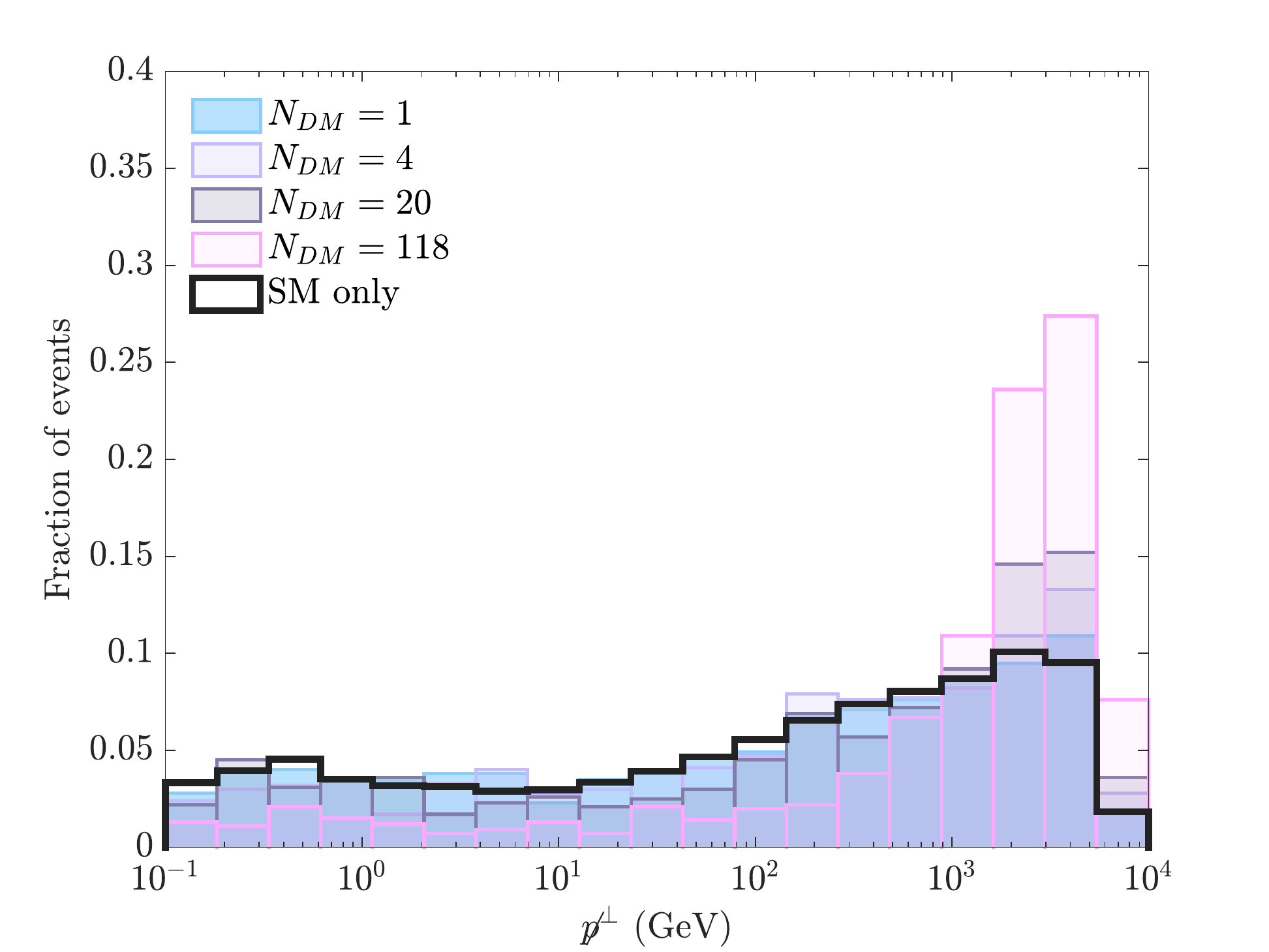} \\
         \includegraphics[trim= 30 0 0 0,clip,width = 0.5\textwidth]{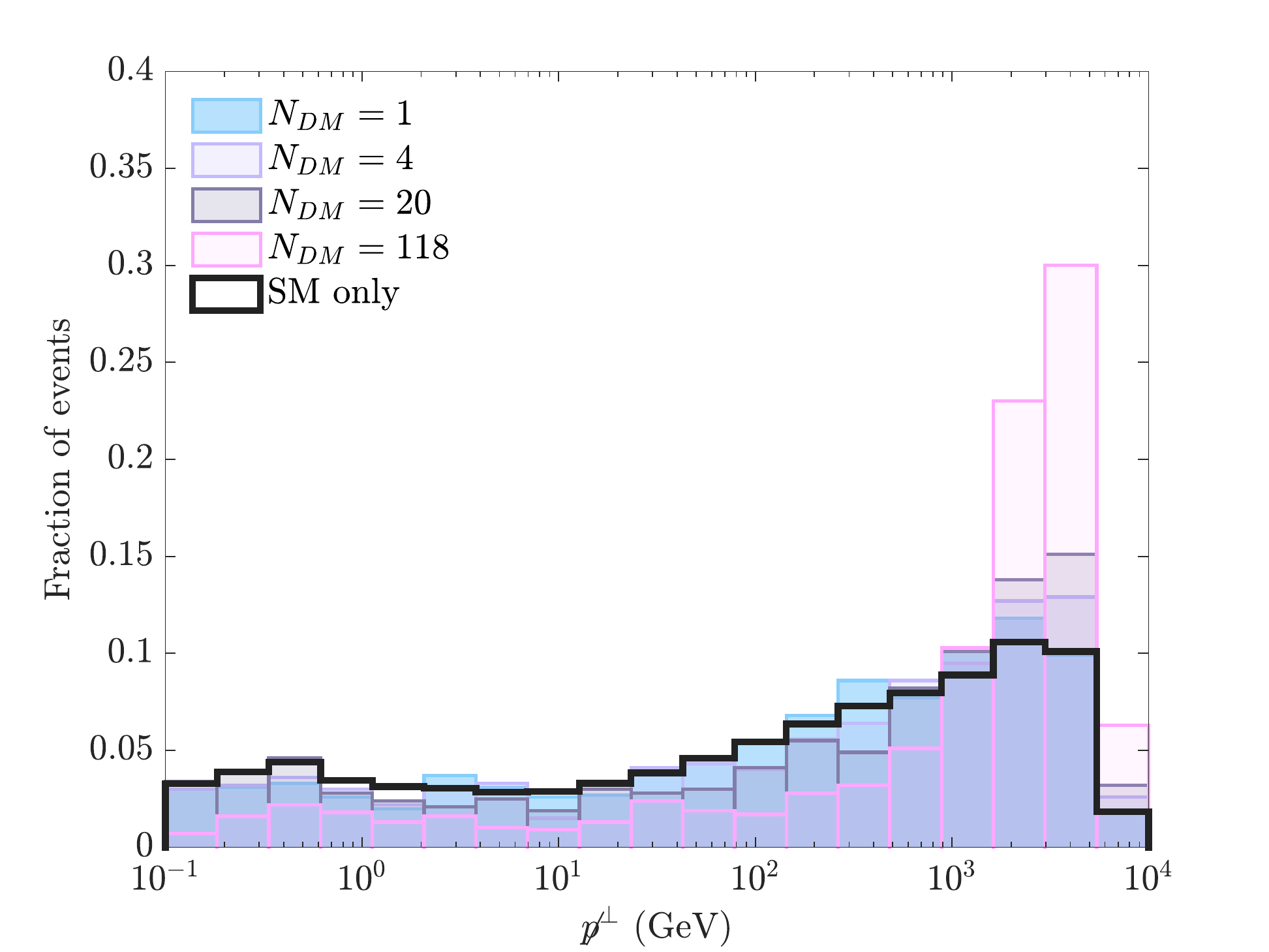} &
      \includegraphics[trim= 30 0 0 0,clip,width = 0.5\textwidth]{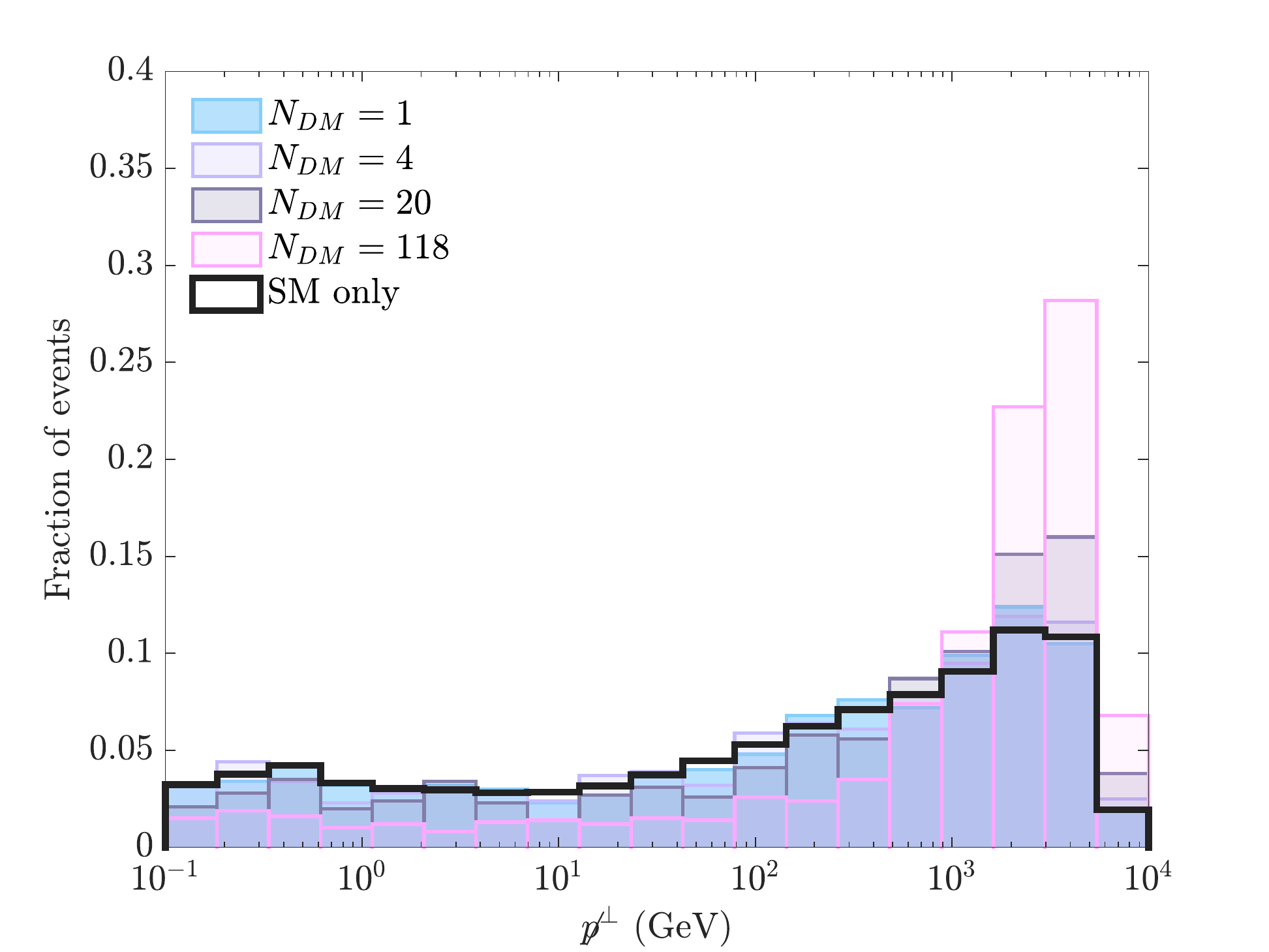} \\
         \includegraphics[trim= 30 0 0 0,clip,width = 0.5\textwidth]{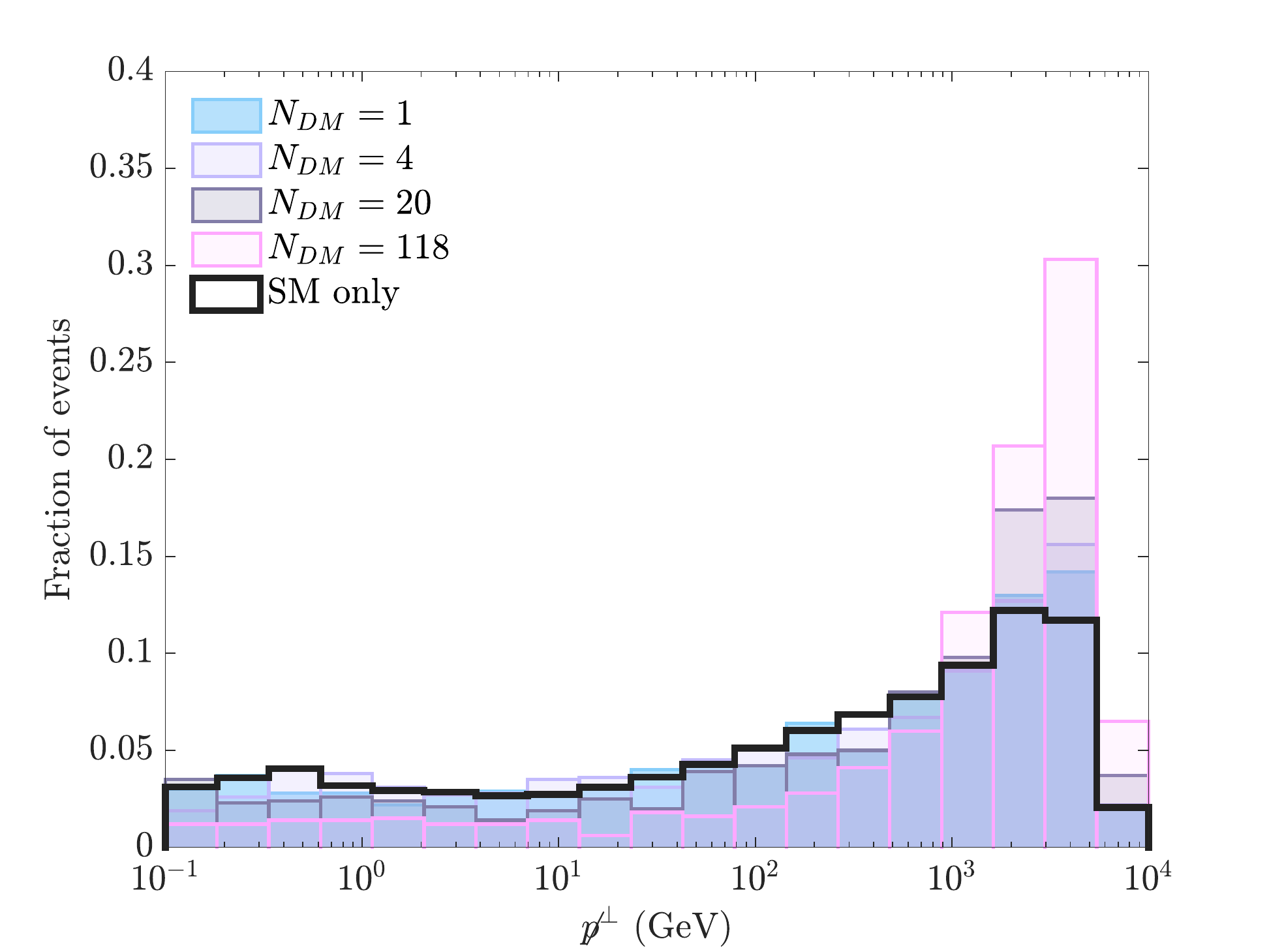} &
      \includegraphics[trim= 30 0 0 0,width = 0.5\textwidth]{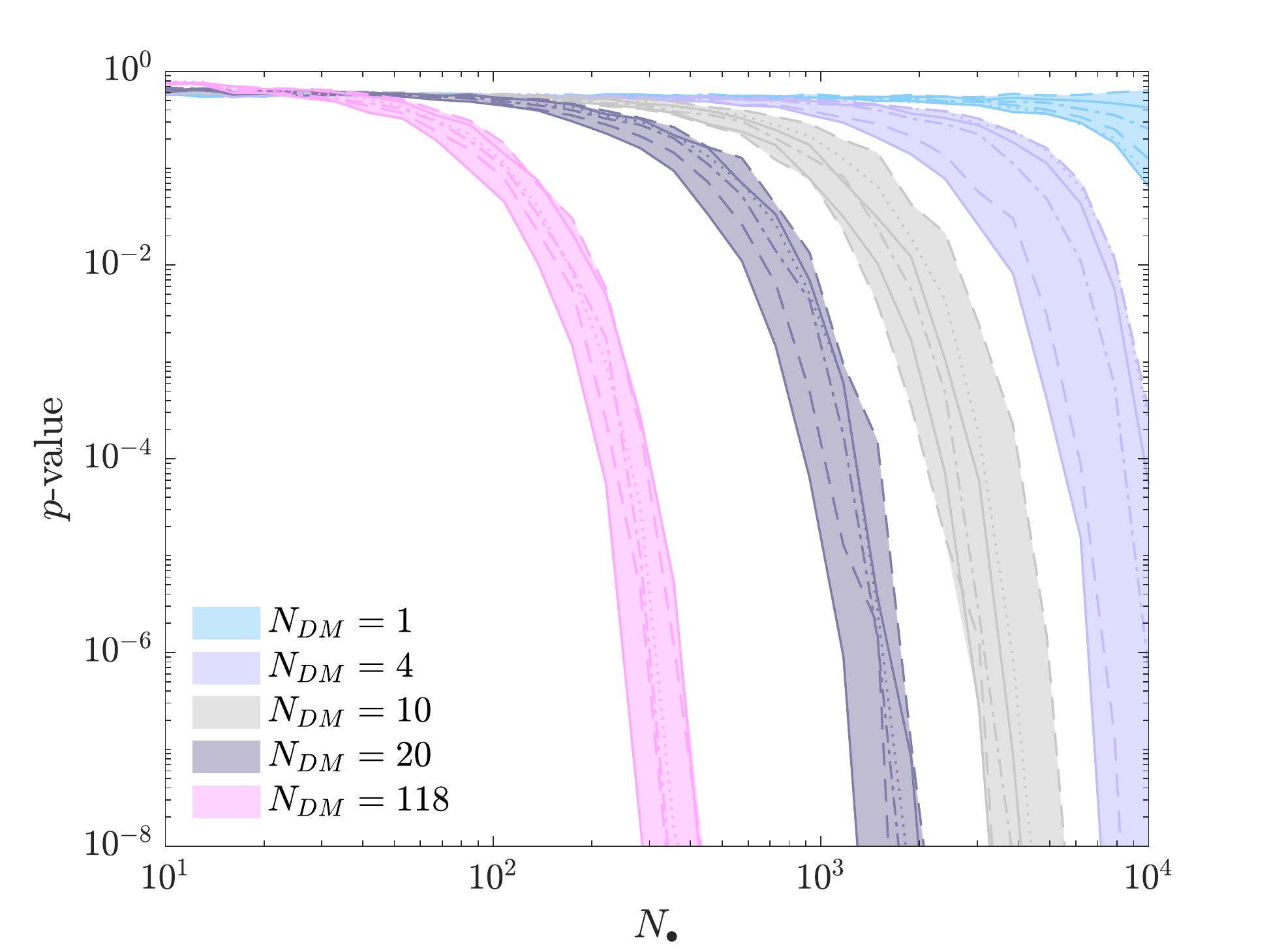} \\
       \end{tabular}
    \caption{Distribution of missing transverse momentum from black hole evaporation for $n = 1,3,4, 5,6$ extra dimensions ($n= 2$ is shown in Fig. 2). Each histogram is made with 1000 black hole creation and evaporation simulations, except for the Standard Model (SM) case, which is made from 10$^6$ black holes. In black we show the SM-only case, and in colour, the case where the dark sector contains 1, 4, 20 and 118 new particles with masses $m_i < M_\bullet$. \textbf{Lower-right:} $p$-value computed with a binned log-likelihood probability function, assuming a chi-squared distributed test statistic, in contrast with the Kolmogorov-Smirnov test shown in Fig. 3.}
    \label{fig:allthefigures}
\end{figure*}

\begin{figure*}
    \begin{tabular}{c c}
   \includegraphics[trim= 30 0 0 0,clip,width = 0.5\textwidth]{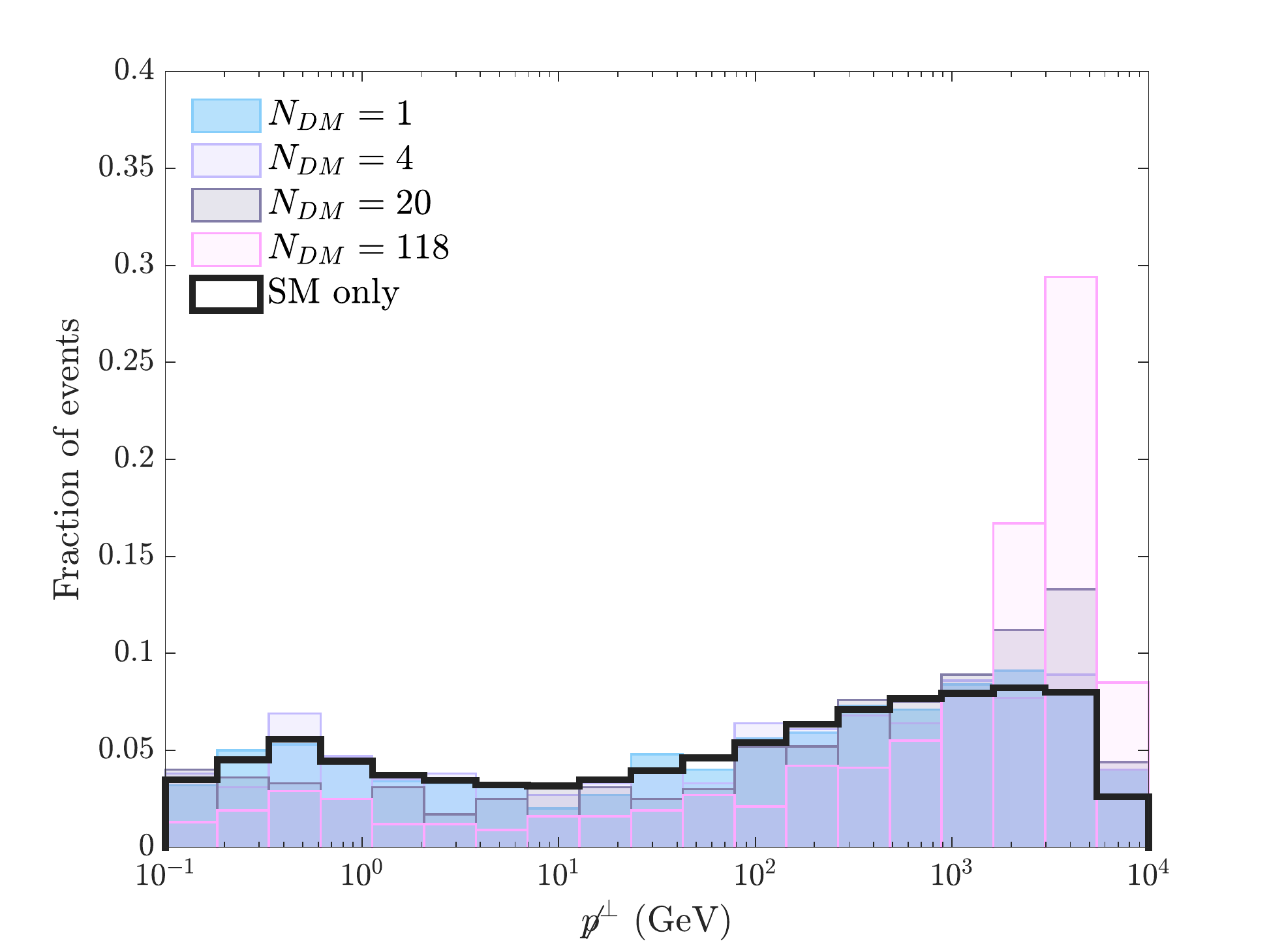} &
      \includegraphics[trim= 30 0 0 0,clip,width = 0.5\textwidth]{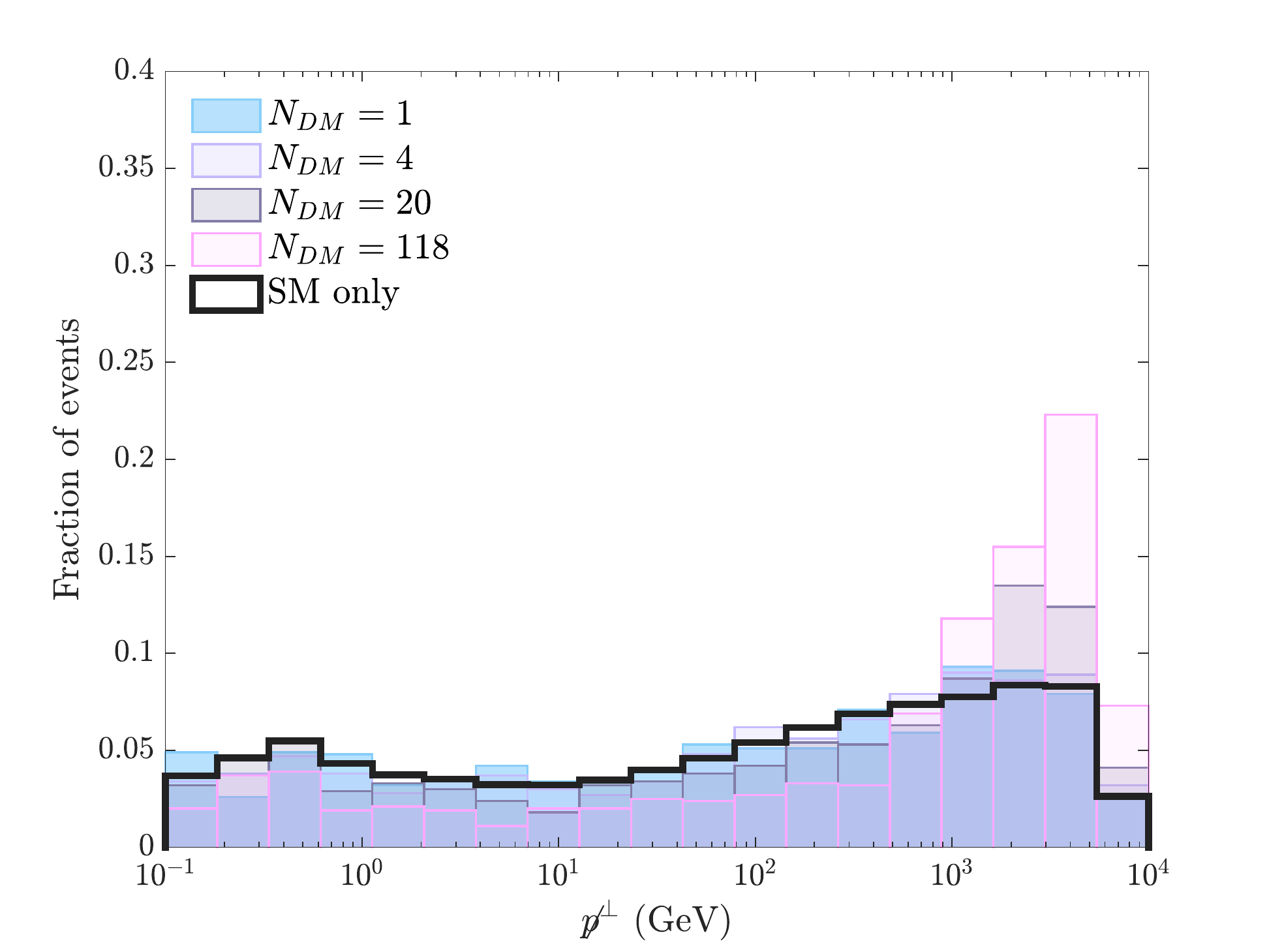} \\
         \includegraphics[trim= 30 0 0 0,clip,width = 0.5\textwidth]{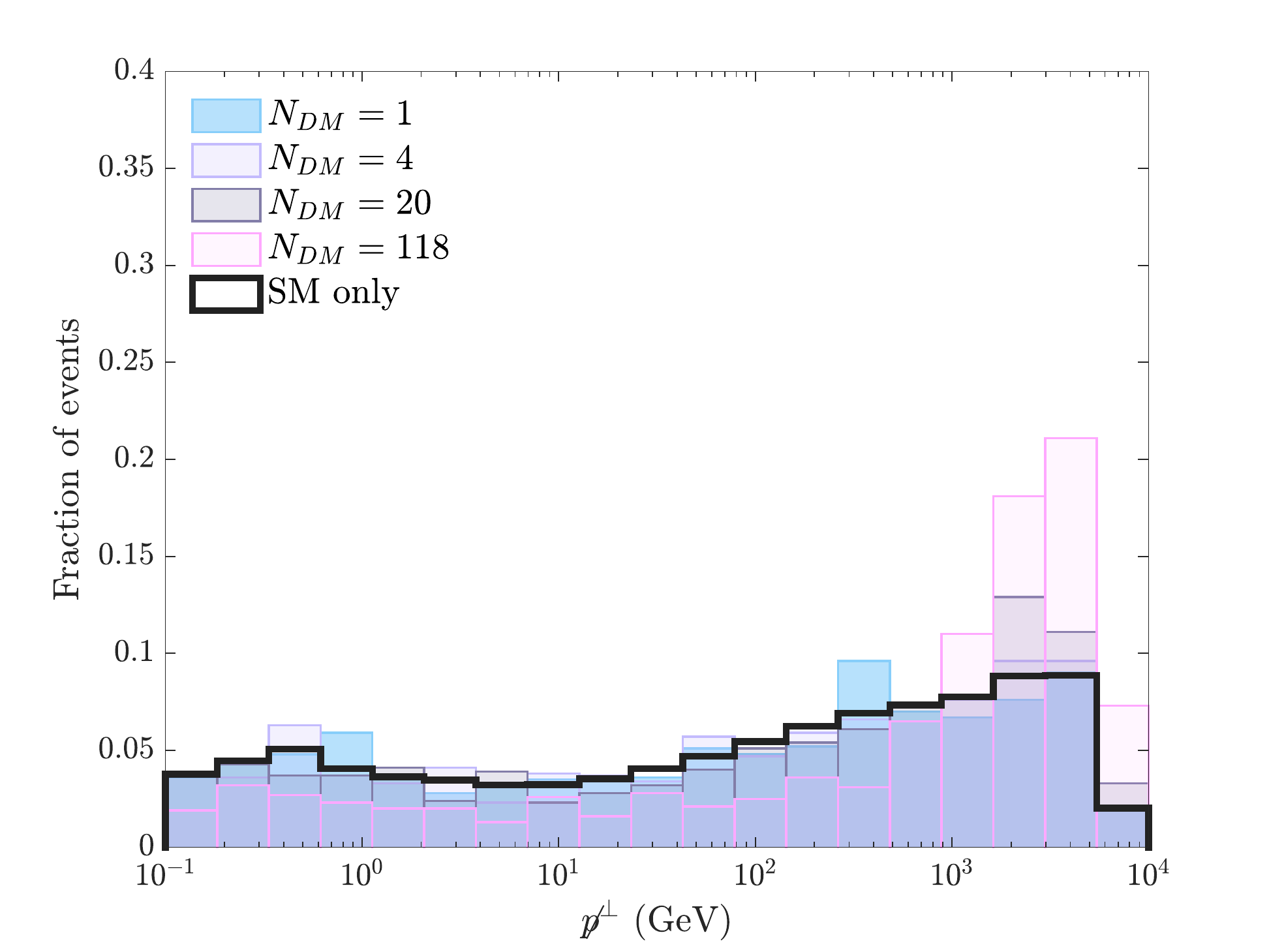} &
      \includegraphics[trim= 30 0 0 0,clip,width = 0.5\textwidth]{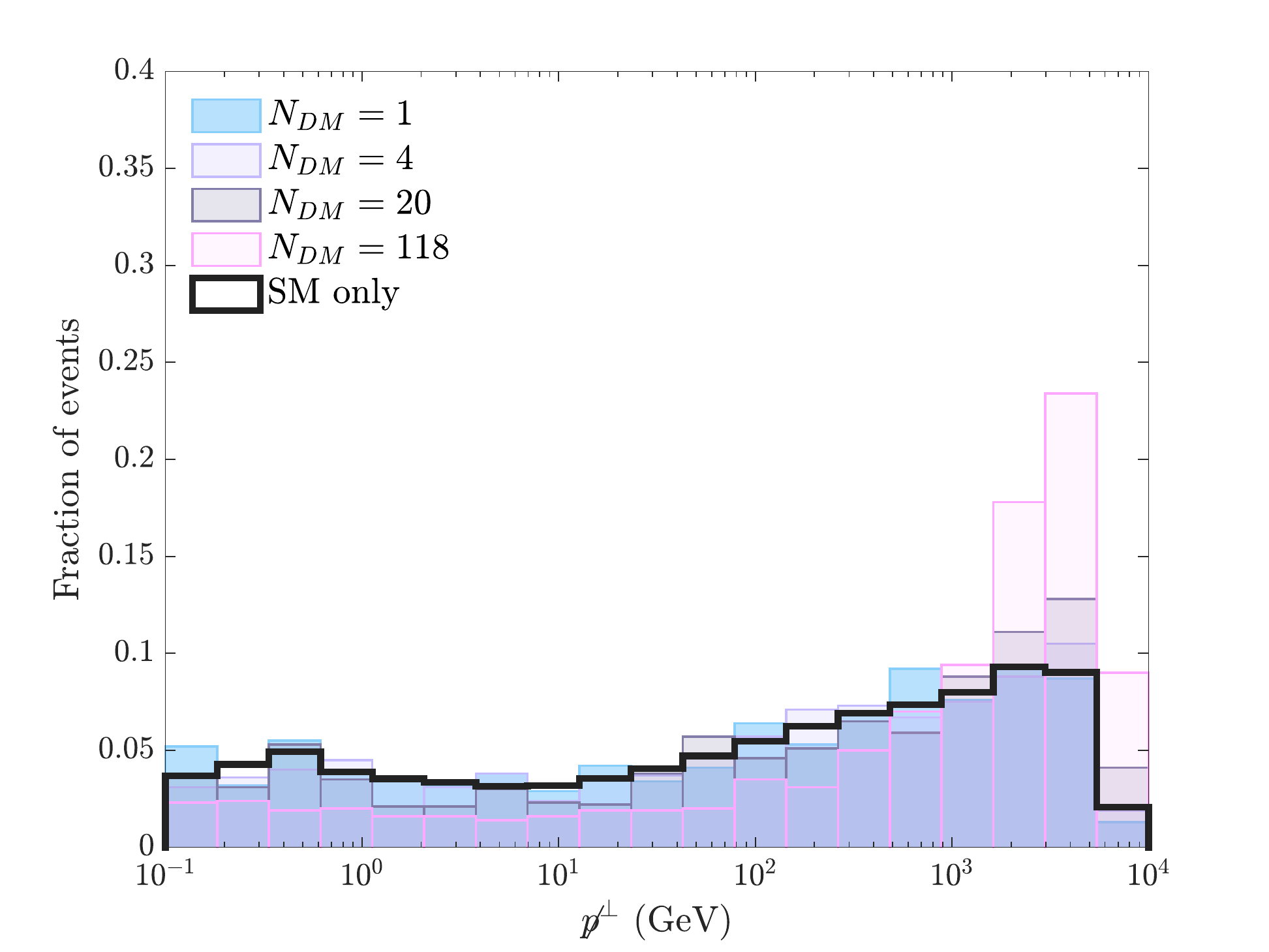} \\
         \includegraphics[trim= 30 0 0 0,clip,width = 0.5\textwidth]{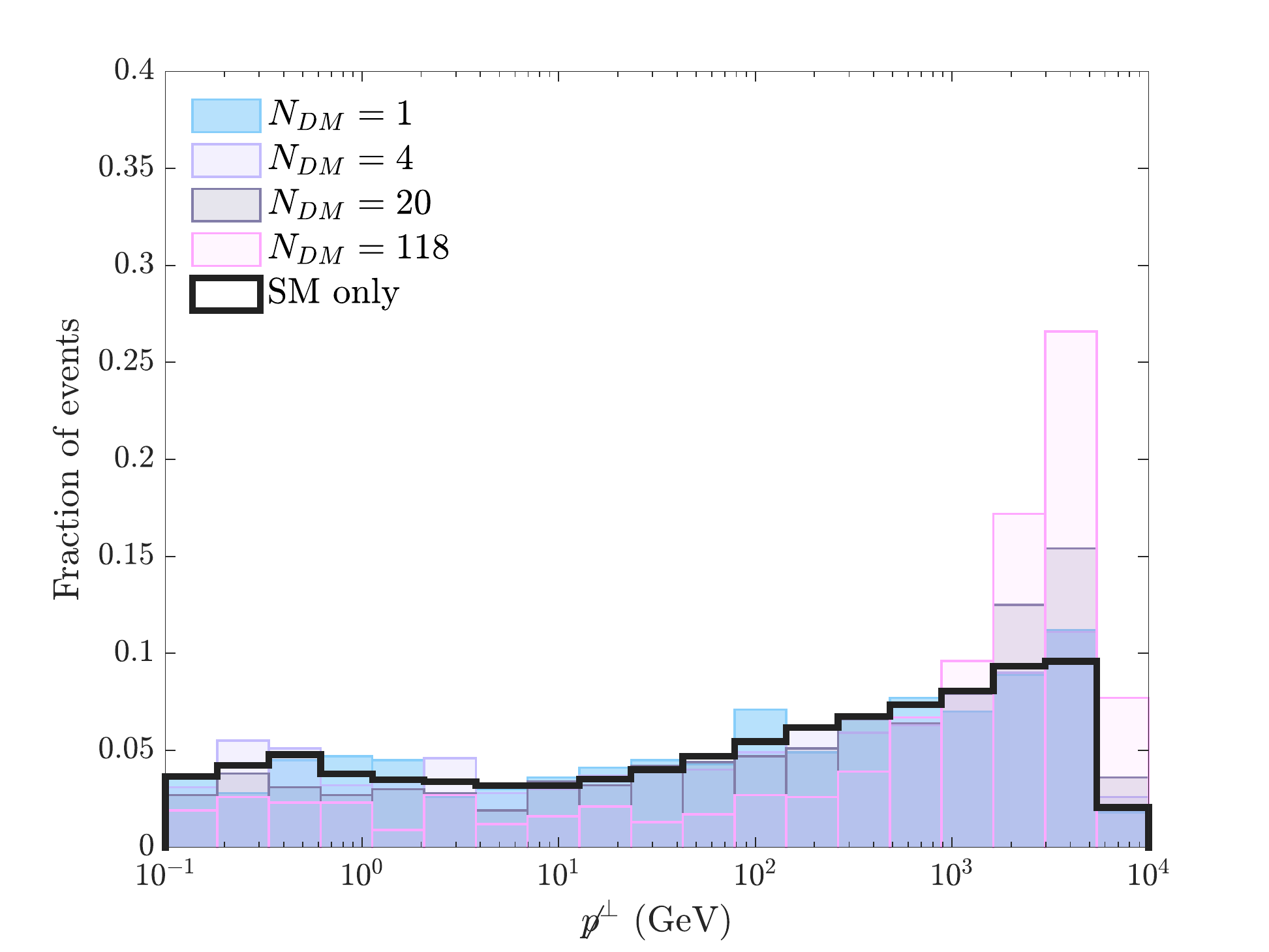} &
      \includegraphics[trim= 30 0 0 0,width = 0.5\textwidth]{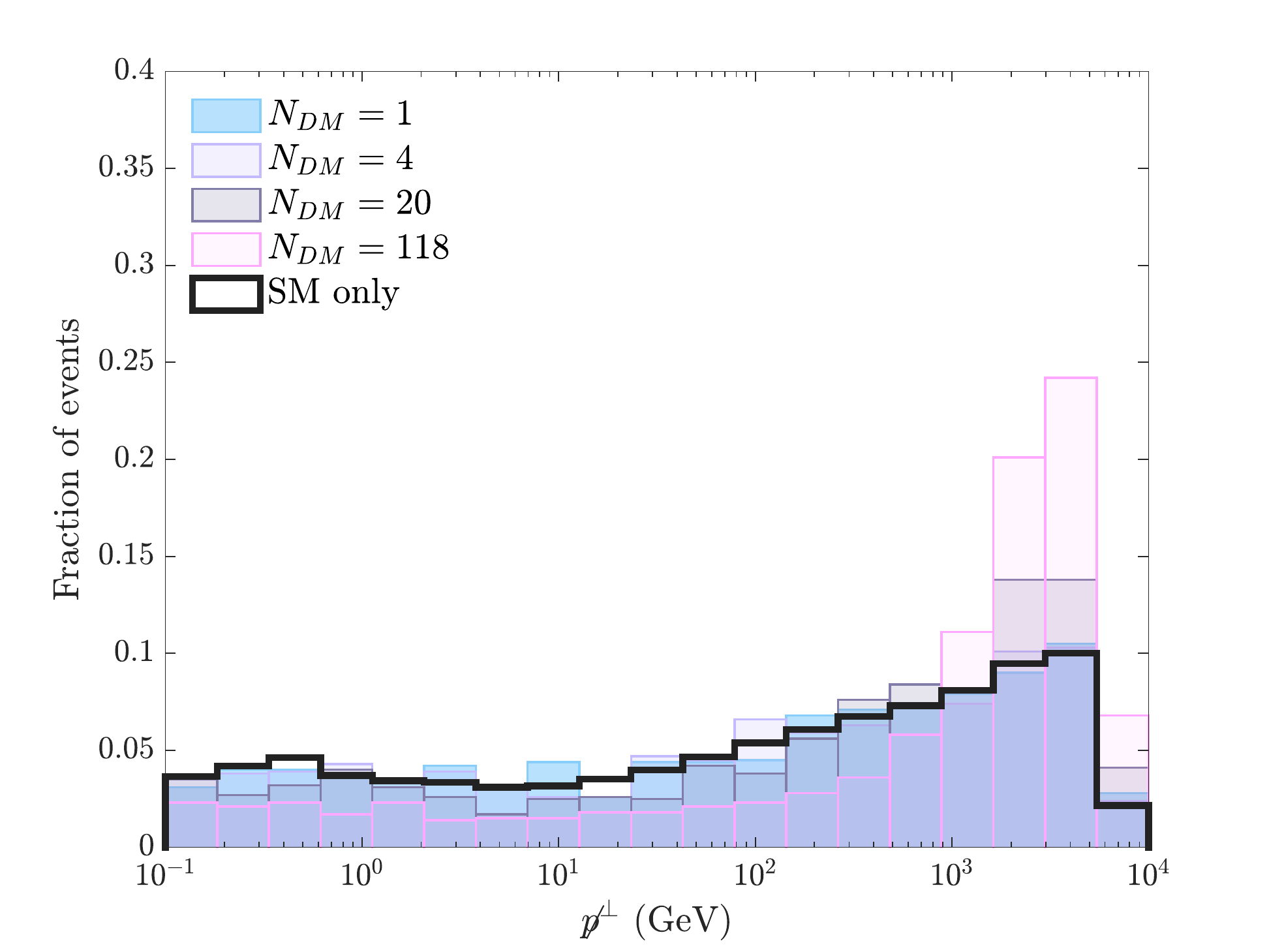} \\
       \end{tabular}
    \caption{Distribution of missing transverse momentum from black hole evaporation for $n = 1$ to $6$ extra dimensions, when BH angular momentum is included -- note that gravitons are not included as greybody factors for bulk tensor emissions and angular momentum are not available. Each histogram  contains 1000 black hole creation and evaporation simulations, except for the Standard Model (SM) case, which is made from 10$^6$ black holes. In black we show the SM-only case, and in colour, the case where the dark sector contains 1, 4 20 and 118 new particles with masses $m_{i} < M_\bullet$.}
    \label{fig:rotating}
\end{figure*}

Here we present missing momentum histograms for $n =$ 1, 3, 4, 5, and 6 extra dimensions in Fig.~\,\ref{fig:allthefigures}. The lower-right panel also shows the $p$-values obtained using a log-likelihood approach:
\begin{equation}
    2\Delta \log L = \sum_i\frac{(N_i - N_{i,SM})^2 }{N_{i,SM}}
    \label{eq:chi2}
\end{equation}{}
$N_i$ is the number of BHs with missing momentum in each bin $i$, and $N_{i,SM}$ is the  standard model-only expectation. The sum runs over 19 logarithmically-spaced bins, and we thus compute $p$-values assuming Eq. \eqref{eq:chi2} to be distributed as a chi-squared with 19 degrees of freedom. Finally, we show for comparison in Fig \ref{fig:rotating}, the corresponding histograms in the case of rotating BHs, noting that bulk graviton emission in this case is not included.

\onecolumngrid

\end{document}